\documentstyle [aps,twocolumn,epsfig]{revtex}
\begin{document}                                                    
\draft 

\twocolumn[

\hsize\textwidth\columnwidth\hsize\csname@twocolumnfalse\endcsname
\begin{center}
{\Large \bf Orbital ordering in undoped manganites using Jahn-Teller
interaction}  \\

\vspace{0.5in}
{\large Sudhakar Yarlagadda and {Manidipa Mitra}$^1$ } \\

\vspace{0.2in}
{Saha Institute of Nuclear Physics, Salt Lake City, Kolkata, India} \\

\vspace{0.2in}
{$^1$S. N. Bose National Centre for Basic Sciences, Salt Lake City, Kolkata, India} \\

\vspace{0.3in}
{Email : sudhakar@cmp.saha.ernet.in }
\end{center}
\maketitle

--------------------------------------------------------------------------------------------------------------------------------------------------------\\
{\small 
{\bf Abstract } :    To understand the orbital ordering of $LaMnO_3$ in 
the ground state, we
study it in the A-type spin antiferromagnetic state.
We calculate the two-dimensional response functions
associated with the Jahn-Teller $Q_2$ and $Q_3$ distortions
and find that they diverge at the wavevector ($\pi$,$\pi$). Furthermore,
the $Q_2$ response function diverges faster which
indicates that perhaps the ordering of the orbitals at low temperatures
is dominated by the $Q_2$ distortion. We calculate the ground state energy
for the cases when only one of the $Q_2$ and $Q_3$ modes
is excited cooperatively and find that the $Q_2$ excited state yields
lower energy.\\ 


 }\\
--------------------------------------------------------------------------------------------------------------------------------------------------------\\
]

\vspace{0.1in}

\noindent
{\bf 1. Introduction}

\vspace{0.1in}

Undoped manganites like $LaMnO_3$ are the parent systems for the colossal
magnetoresistive materials. It is well known that orbital ordering occurs
around 780 K resulting in a C-type orbital structure with two kinds
of orbitals alternating on adjacent sites in the $xy$ plane while
like orbitals are stacked in the $z$ direction [1]. 
 As the temperature is further
lowered to 140 K an A-type spin antiferromagnetic order sets in wherein
the spins are ferromagnetically aligned in the $xy$ plane with the spin
coupling in the $z$ direction being antiferromagnetic [2].
To explain the
observed order several studies have been reported. These studies fall
into two classes, with one class based on Jahn-Teller (JT)
interaction [3--6]
being the main cause while the other class corresponds to Coulombic
interaction [7--10] 
being the dominant cause for the observed order. However,
it must be pointed out that the electron-phonon based studies do not point to
a transparent mechanism.
 
The purpose of the present paper is to
use the  framework developed earlier by one of us  
(Ref. [11]) and analyze
the orbital ordering by using a generalized Peierls instability approach.
However, as compared to the one-dimensional Peierls charge density wave (CDW)
approach, our higher dimensional orbital density wave (ODW) analysis is
more complicated on account of there being 
two $e_g$ orbitals (with inter-orbital hopping) and two response functions
corresponding to the JT $Q_2$ and $Q_3$ distortions.
We begin by assuming that the A-type antiferromagnetic ordering 
has already set in. Thus, on account of strong Hund's coupling,
the transport is restricted to spin polarized electrons in two dimensions only.
 We assume that the $Mn$ sites are fixed and that only the oxygen
atoms between the $Mn$ sites can move along the axis joining the neighboring
$Mn$ sites.
We find that the non-interacting response functions
$\chi_{2,3}$ corresponding to the $Q_{2,3}$ modes diverge at
wavevector $\vec{Q}=(\pi, \pi)$. Thus one can expect local minima in
energy for states corresponding to ODW being excited at
wavevector $\vec{Q}$ by the cooperative $Q_{2}$ and $Q_{3}$ modes. 
We find that $\chi_2 ( \pi,\pi)$ diverges faster than
$\chi_3 ( \pi,\pi)$ and as expected the energy for the $Q_2$ related
ODW state has lower energy than the $Q_3$ related ODW state. 

\vspace{0.18in}

\noindent
{\bf 2. Mean-field approach to ODW} 

\vspace{0.1in}

We will now consider manganite systems with two $e_g$ orbitals per
site and ignore spin. The Hamiltonian consists of 
the kinetic term, the ionic term, and the electron-ion interaction term.
The kinetic term in momentum space is given by 
\begin{equation}
H_1 = \sum _{\vec{p}}{\bf B}^{\dagger}_{\vec{p}}\cdot{\bf T} 
\cdot {\bf B}_{\vec{p}} ,
\end{equation}
where $ {\bf B}^{\dagger}_{\vec{p}} \equiv
(b^{\dagger}_{1\vec p} , b^{\dagger}_{2 \vec p})$ with
$b_1$  and $b_2$ 
corresponding to the destruction operators
for electrons with the orthonormal wavefunctions
 $\psi_{x^2-y^2}$ and $\psi_{3z^2 -r^2}$ respectively.
Furthermore, ${\bf T}$ is a hermitian matrix with
 ${\bf T}_{1,1}=-1.5 t[\cos p_x + \cos p_y]$, 
${\bf T}_{2,2}=-0.5 t[\cos p_x + \cos p_y]$, and
 ${\bf T}_{1,2}=0.5 \sqrt{3} t[\cos p_x - \cos p_y ]$.
The eigen values of the kinetic energy are given by
$\lambda_{n}^{\vec{p}} = -\cos p_x - \cos p_y - (-1)^n \sqrt{\cos^2 p_x 
+ \cos^2 p_y  - \cos p_x  \cos p_y }$ with $n=1,2$.
The Fermi sea corresponding to the lower eigen energy value 
$\lambda_{2}^{\vec{p}}$ is given by the union of the region 
$-\pi/2 \le  k_x \le \pi/2$ (with all values of $k_y$ allowed) and
the region   
$-\pi/2 \le  k_y \le \pi/2$ (with all values of $k_x$ allowed).  
Whereas the Fermi sea corresponding to the higher eigen energy value 
$\lambda_{1}^{\vec{p}}$
is given by the intersection of the region 
$-\pi/2 \le  k_x \le \pi/2$ and the region
$-\pi/2 \le  k_y \le \pi/2$. Since the number of electrons is equal to
the number of sites, the total area
occupied by both Fermi seas is equal to the area of the
Brillouin zone ($4 \pi^2$). Furthermore, the Fermi surface corresponds
to $\lambda_{n}^{\vec{p}} = 0$.  

Using  mean-field approximation we get, 
 after averaging the Hamiltonian over phonon coordinates,
 the following effective Hamiltonian
(see Ref. [11] for details):
 \begin{eqnarray} 
\bar{H}= &&
  H_{1} 
  -
 2 \frac{A^2}{\omega_0} \sum_{ j } 
\left [ (b^{\dagger}_{1j} b_{2j} +
b^{\dagger}_{2j} b_{1j})
\langle b^{\dagger}_{1j} b_{2j} +
b^{\dagger}_{2j} b_{1j} \rangle \right . 
\nonumber \\
&&
 +
\left . (b^{\dagger}_{1j} b_{1j} - 
b^{\dagger}_{2j} b_{2j}) 
\langle
b^{\dagger}_{1j} b_{1j} - 
b^{\dagger}_{2j} b_{2j}
 \rangle \right ]
 \nonumber \\                           
  && 
  +
  \frac{A^2}{\omega_0} \sum_{ j }
\left [ \langle
b^{\dagger}_{1j} b_{2j} +
b^{\dagger}_{2j} b_{1j}
 \rangle ^2 +
\langle
b^{\dagger}_{1j} b_{1j} - 
b^{\dagger}_{2j} b_{2j}
 \rangle ^2 \right ]  ,
\label{2dodw}
 \end{eqnarray} 
where $<..>$ implies averaging over the phonon coordinates,
$A \sqrt{2 M \omega_0}$ is the electron-JT coupling constant, 
$2 A \langle b^{\dagger}_{1j} b_{2j} + 
b^{\dagger}_{2j} b_{1j} \rangle =- \omega_0 \sqrt{2 M \omega_0} \langle
Q_{2j} \rangle $, and
$2 A \langle b^{\dagger}_{1j} b_{1j} - 
b^{\dagger}_{2j} b_{2j} \rangle =- \omega_0 \sqrt{2 M \omega_0} \langle
Q_{3j} \rangle $.

We will now determine the wavevector for long range cooperative ordering
of the $Q_{2,3}$ modes. For this we must figure out the values
 of $q$ that make the susceptibilities $\chi_{2,3} (q,q)$
diverge. The expressions for 
 $\chi_{2,3} (\vec{q})$ are given as
\begin{eqnarray} 
&& \chi_{l} (\vec{q})= 
-2 \sum_{\vec{k},n}\frac{\langle c^{n\dagger}_{\vec{k}}
c^{n}_{\vec{k}}\rangle}{ 
\lambda^{\vec{k}+\vec{q}}_{n} -\lambda^{\vec{k}}_{n}}
\sin^2 \left [ \frac{\theta_{\vec{k}+\vec{q}}
 +\theta_{\vec{k}}+ l \pi }{2}
\right ]
\nonumber \\
&&
+2 \sum_{\vec{k}}\frac
{\langle c^{1\dagger}_{\vec{k}+\vec{q}} 
c^{1}_{\vec{k}+\vec{q}}\rangle -
\langle c^{2\dagger}_{\vec{k}} c^{2}_{\vec{k}}\rangle}
{ \lambda^{\vec{k}+\vec{q}}_{1} -\lambda^{\vec{k}}_{2}}
\cos^2 \left [ \frac{\theta_{\vec{k}+\vec{q}}
 +\theta_{\vec{k}}+l \pi}{2}
\right ] ,
 \end{eqnarray} 
where $( c^{1\dagger}_{\vec{k}}, c^{2\dagger}_{\vec{k}})=
(b^{\dagger}_{1 \vec{k}}, b^{\dagger}_{2 \vec{k}}) \cdot {\bf M}$,
 ${\bf M}$ is the diagonalizing matrix for
the kinetic matrix ${\bf T}$ with 
${\bf M}_{1,1}=\sin(\theta_{\vec{k}}/2)$,
${\bf M}_{2,2}=-\sin(\theta_{\vec{k}}/2)$, and
${\bf M}_{1,2}=\cos(\theta_{\vec{k}}/2)$.
At 0 K the second term on the right hand side  of both $\chi_2$ and
$\chi_3$ produces a divergence at $\vec{Q}$ because of the fact that 
$ \lambda^{\vec{k}+\vec{Q}}_{1} = -\lambda^{\vec{k}}_{2}$ and that
the Fermi energy is zero. Furthermore the ratio of $\chi_2 / \chi_3$
at 0 K and wavevector $\vec{Q}$ is expected to be 3.

\vspace{0.18in}

\noindent
{\bf 3.  Results and discussion} 

\vspace{0.1in}

In Fig. 1 we plot $\chi_{2,3} (\vec{Q})$ as a function of
temperature (with hopping term $t$ set equal to 0.1 eV)
and notice that they diverge as $T \rightarrow 0$
with $\chi_2$ diverging faster than $\chi_3$.
In Fig. 2 we plot $\chi_{2,3} (q,q)$ as a function of $q$ 
(with $0 \le q \le \pi$)
at fixed temperature T = 0.01 K and hopping term t = 0.1 eV
 and find that they both peak sharply 
close to $\vec{Q}$ with $\chi_2$ peaking faster. Then based on the fact that
the phonon mode goes soft at wavevector $\vec{Q}$ we compute the
ground state energy when only either $Q_2$ mode or $Q_3$ mode gets excited
cooperatively in the system. The order parameters are given by
$\langle b^{\dagger}_{1j} b_{2j} + 
b^{\dagger}_{2j} b_{1j} \rangle =
 c_2 \cos (\vec{Q} \cdot \vec{R_j})$  and
$\langle b^{\dagger}_{1j} b_{1j} - 
b^{\dagger}_{2j} b_{2j} \rangle =
 c_3 \cos (\vec{Q} \cdot \vec{R_j})$ with $-1 \le c_{2,3} \le 1$ 

\begin{figure}
\resizebox*{3.0in}{2.7in}{\rotatebox{0}{\includegraphics{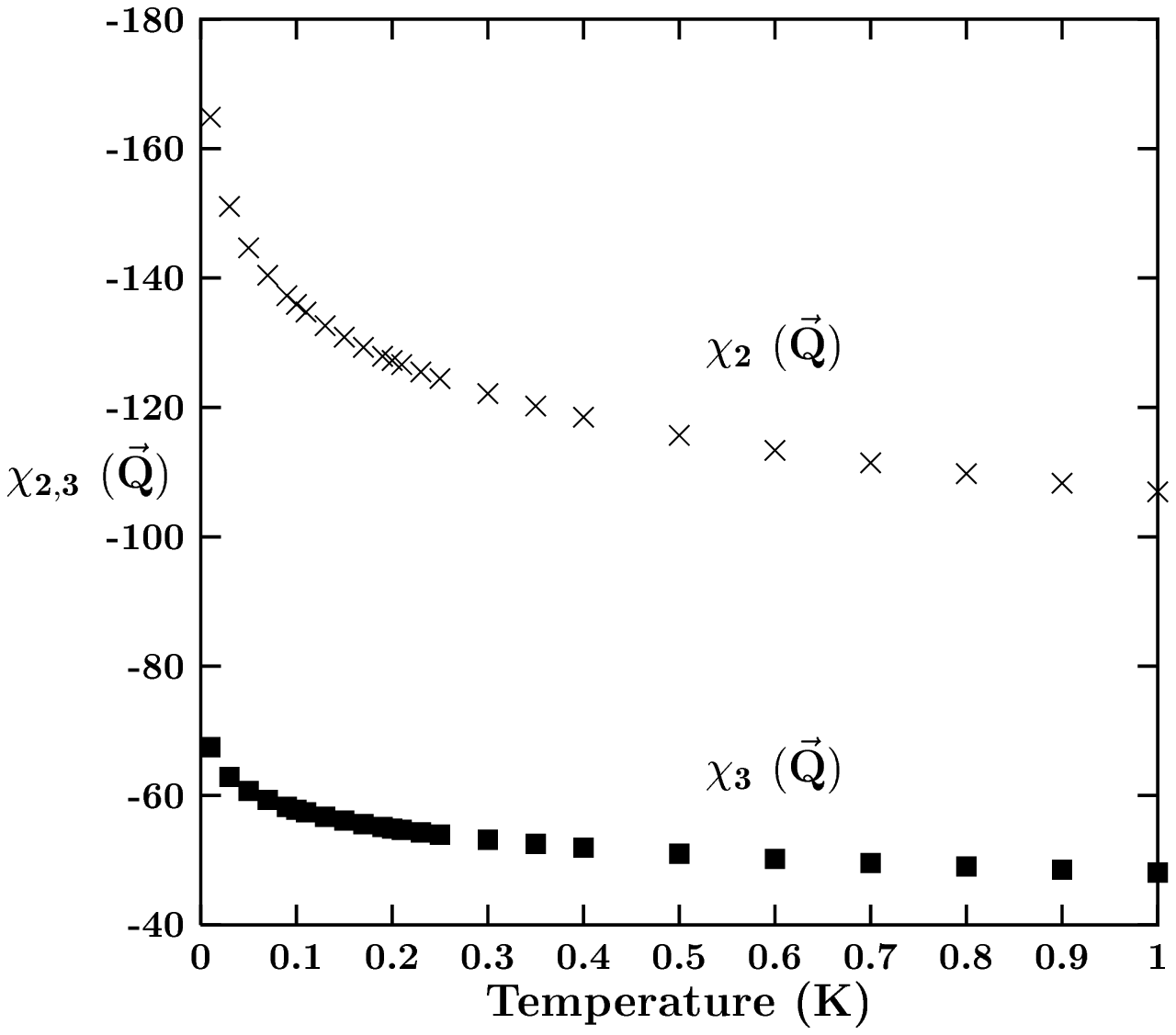}}}
\label{scaling}
\end{figure}
\noindent
{\small{\bf Figure 1. }Plot of $\chi_{2,3} (\pi,\pi)$ as a function of
temperature at $t=0.1 ~ eV$.}
\vspace*{0.5cm}
\begin{figure}
\resizebox*{3.0in}{2.7in}{\rotatebox{0}{\includegraphics{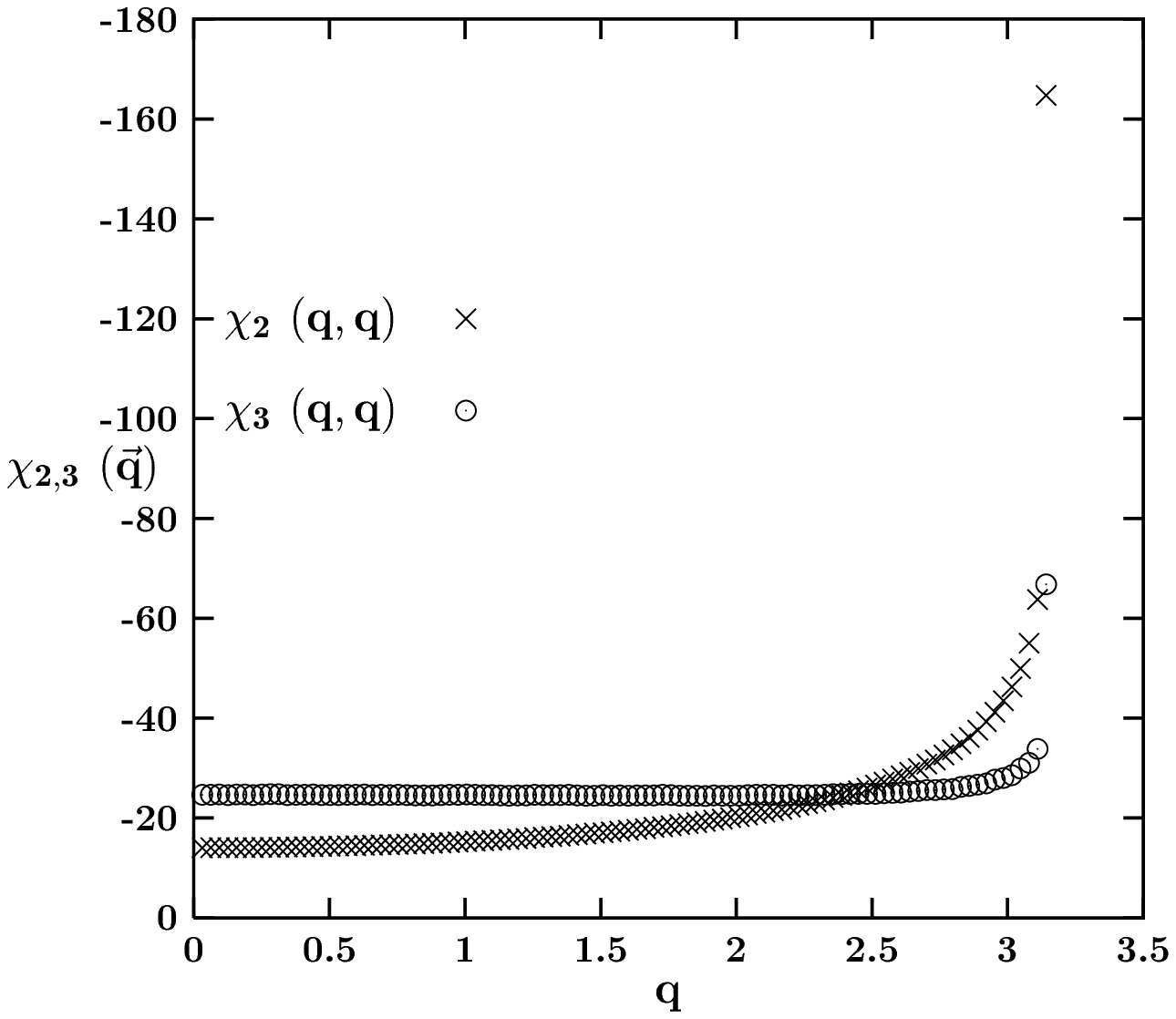}}}
\label{fig2}
\end{figure}

\noindent
{\small {\bf Figure 2.}  Plot of $\chi_{2,3} (q,q)$ for different values 
of ($0 ~\le q ~\le ~\pi $) at T = 0.01 K and $t = 0.1 ~eV$.} \\ 

\noindent
and $\vec R_j$ being the position vector. Here it should be 
pointed out that the order parameter
$\langle b^{\dagger}_{1j} b_{2j} + 
b^{\dagger}_{2j} b_{1j} \rangle$
corresponds to the density difference of electrons in the
two orbitals
 $\psi_X \equiv (\psi_{x^2-y^2} - \psi_{3z^2 -r^2})/\sqrt{2}$ 
 and $ \psi_Y \equiv -(\psi_{x^2-y^2} + \psi_{3z^2 -r^2})/\sqrt{2}$ (see  
Ref. [5]).
From the symmetry of the $\psi_X$ and $\psi_Y$ orbitals it
follows that at each site the total charge is unity. 
The unit cell needed to
compute the ground state energy
consists of two adjacent sites with the Brillouin zone being given by
$-\pi ~\le (k_x + k_y) ~\le \pi $
and $-\pi ~\le (k_x - k_y) ~\le \pi $. We
diagonalize 
a $4 \times 4$ matrix at each momentum and integrate the lowest
two eigen energies over the Brillouin zone to obtain the ground state
energy.
The results of 
\begin{figure}
\resizebox*{3.0in}{2.7in}{\rotatebox{0}{\includegraphics{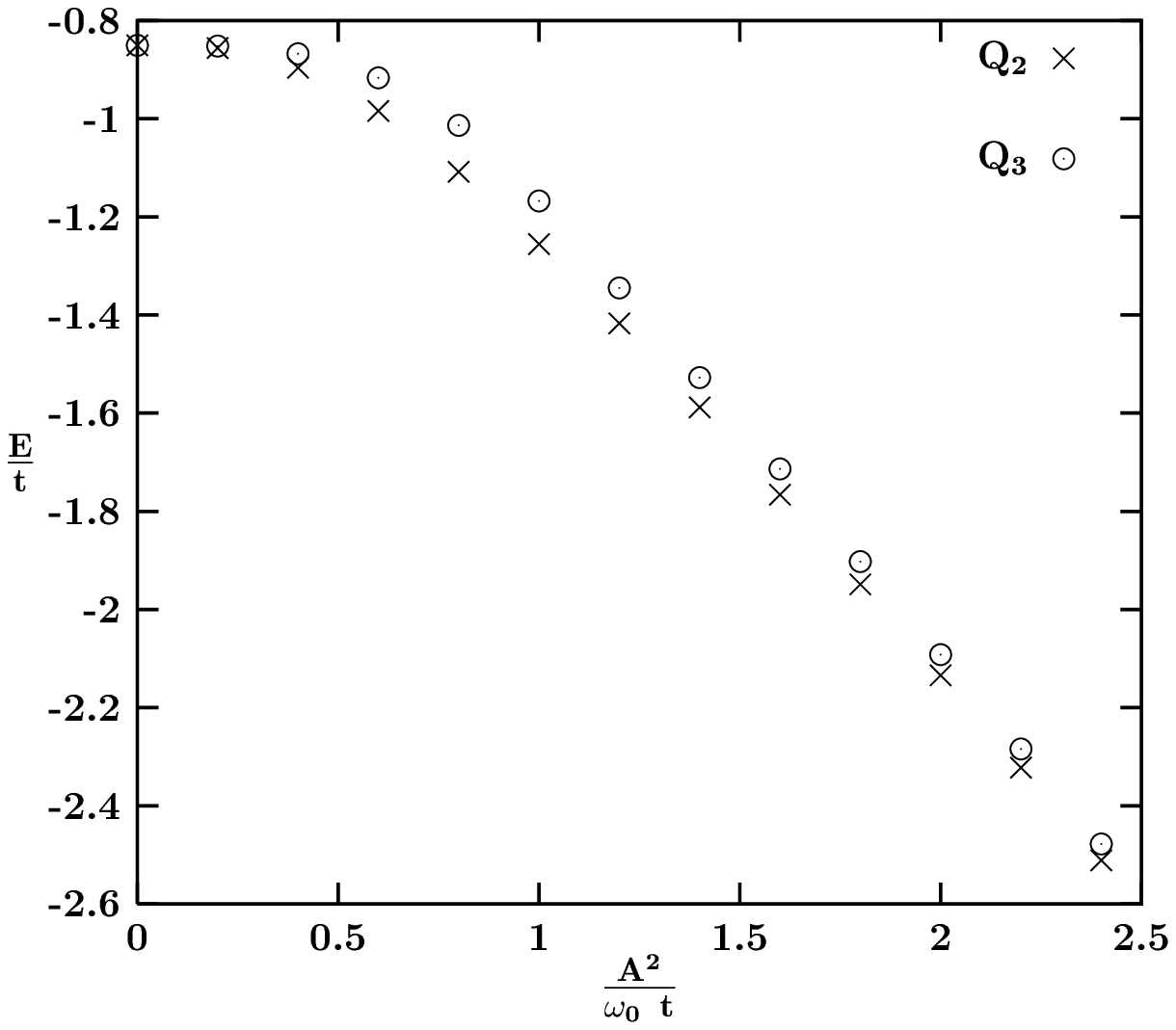}}}
\label{fig3}
\end{figure}
\noindent
{\small 
{\bf Figure 3.}  Dependence on dimensionless polaronic energy 
($A^2 /\omega_0 t$) of dimensionless ground state energy per site ($ E/t $)  
for cooperative $Q_2$ and $Q_3$ modes.}\\

\begin{figure}
\resizebox*{3.0in}{2.7in}{\rotatebox{0}{\includegraphics{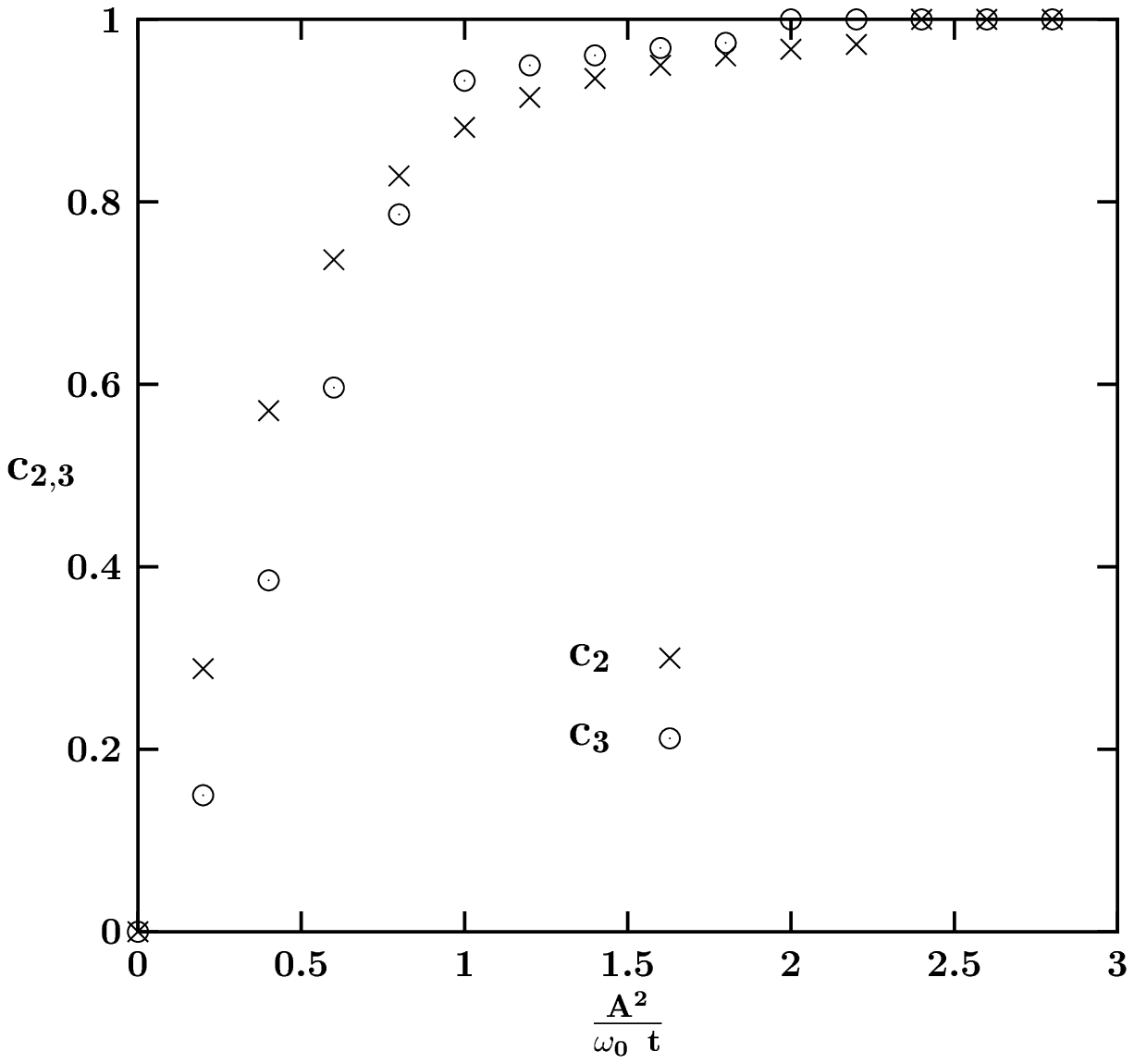}}}
\label{fig4}
\end{figure}
\noindent
{\small {\bf Figure 4.} Variation of coefficients $c_{2,3}$ of
ODW order parameters for $Q_2$ and $Q_3$ distortions as a function
of the dimensionless polaronic energy 
 ($A^2 /\omega_0 t$).}\\

\noindent
our calculations are shown in Figs. 3 and 4. From Fig. 3
we see that the ground state energy corresponds to the $Q_2$ mode
with the difference in energy between the $Q_2$ only state and the $Q_3$
only state peaking at intermediate values of the dimensionless polaronic energy
($A^2/\omega_0 t$).
For zero values and infinite values of
the polaronic energy both modes yield the same energy
because zero value implies no phononic coupling effect while infinite value
corresponds to localized polarons. 
Thus for
large values of the polaronic energy,
the ground  state energy is only slightly smaller than the polaronic energy. 
Furthermore, from Fig. 4 we also see that as the polaronic energy increases
the values of $c_{2,3}$ increase and become unity around
 $A^2/(\omega_0 t) \sim 2$ implying that for the $Q_3$ ($Q_2$) mode
$\psi_{x^2-y^2}$ ($\psi_X$) orbital is occupied
fully at one site with the $\psi_{3z^2 -r^2}$ ($\psi_Y$)
 orbital being fully occupied at the adjacent sites.

In conclusion, we have studied orbital ordering for the ground state
of the undoped manganite systems. We find that the two-dimensional
orbital ordering, in the ferromagnetic planes of the observed
A-type antiferromagnetic
state, is governed by the wavevector $\vec{Q}$ with $Q_2$ JT mode
being cooperatively excited in the system. \\

\vspace{0.5in}

\noindent
{\bf References } \\
\begin{itemize}
\item[[1]] Y. Murakami, J. P. Hill, D. Gibbs, M. Blume, I. Koyama, 
M. Tanaka, H. Kawata, T. Arima, Y. Tokura, K. Hirota, and Y. Endoh, 
   Phys. Rev. Lett.  {\bf 81},                      
   582 (1998).                           
\item[[2]] E. O. Wollan and W. C. Koehler,
Phys. Rev. {\bf 100}, 545 (1955). 
\item[[3]] A. J. Millis, 
   Phys. Rev. B  {\bf 53},                      
   8434 (1996).                          
\item[[4]] T. Hotta,  S. Yunoki, M. Mayr, and E. Dagotto,
   Phys. Rev. B  {\bf 60},                      
  R15009 (1999).                          
\item[[5]] P. B. Allen and V. Perebeinos, Phys. Rev. B {\bf 60},
 10747 (1999). 
\item[[6]] Z. Popovic and S. Satpathy,
   Phys. Rev. Lett.  {\bf 84},                      
   1603 (2000).                           
\item[[7]] K. I. Kugel and D. I.  Khomskii,
   Sov. Phys. JETP  {\bf 37},                      
   725 (1973).                          
\item[[8]] S. Ishihara, J. Inoue, and S. Maekawa, Phys. Rev. B {\bf 55},
8280 (1997).  
\item[[9]] L. Sheng and D. N. Sheng, Int. J. Mod. Phys. {\bf B 13}, 1397 (1999).
\item[[10]] S. Okamoto, S. Ishihara, and S. Maekawa,
 Phys. Rev. B {\bf 65},
 144403 (2002). 
\item[[11]] S. Yarlagadda,  
Int. J. Mod. Phys. B {\bf 15}, 3529 (2001). 
\end{itemize}
\end{document}